\documentclass{amsart}

\usepackage{amsbsy,amssymb,amscd,amsfonts,latexsym,amstext,delarray,
amsmath,graphicx}
\input xypic
\usepackage{color}

\numberwithin{equation}{section}

\def\bF{{\mathbb F}}

\def\bP{{\mathbb P}}

\def\bS{{\mathbb S}}
\def\bT{{\mathbb T}}

\def\C{{\mathbb C}}
\def\F{{\mathbb F}}
\renewcommand{\H}{{\mathbb H}}

\renewcommand{\P}{{\mathbb P}}
\def\Q{{\mathbb Q}}
\def\Z{{\mathbb Z}}
\def\R{{\mathbb R}}

\def\cA{{\mathcal A}}
\def\cB{{\mathcal B}}
\def\cC{{\mathcal C}}

\def\cH{{\mathcal H}}

\def\cS{{\mathcal S}}

\def\cV{{\mathcal V}}

\def\Tr{{\rm Tr}}
\def\Spec{{\rm Spec}}
\def\SL{{\rm SL}}
\def\GL{{\rm GL}}

\def\fs{{\mathfrak{s}}}
\def\fa{{\mathfrak{a}}}

\def\cancel#1#2{\ooalign{$\hfil#1\mkern1mu/\hfil$\crcr$#1#2$}}
\def\Dirac{\mathpalette\cancel D}
\def\dirac{\mathpalette\cancel\partial}

\def\cutint{{\int \!\!\!\!\!\! -}}

\title{Noncommutative Mixmaster Cosmologies}
\author{Christopher Estrada and Matilde Marcolli}
\address{Mathematics Department, Mail Code 253-37, Caltech, 1200 E.~California Blvd. Pasadena, CA 91125, USA}
\email{c.estrada@caltech.edu}
\email{matilde@caltech.edu}

\begin{document}

\maketitle

\begin{abstract} In this paper we investigate a variant of the classical
mixmaster universe model of anisotropic cosmology, where the spatial
sections are noncommutative $3$-tori. We consider ways in which the
discrete dynamical system describing the mixmaster dynamics can be
extended to act on the noncommutative torus moduli, and how the
resulting dynamics differs from the classical one, for example, in
the appearance of exotic smooth structures. We discuss
properties of the spectral action, focussing on how the slow-roll 
inflation potential determined by the spectral action affects the
mixmaster dynamics. We relate the model to other recent results
on spectral action computation and we identify other physical
contexts in which this model may be relevant.
\end{abstract}

\tableofcontents

\section{Introduction}

Noncommutative geometry has been used extensively as a method for the
construction of models of particle physics, cosmology, and gravity coupled 
to matter. The point of view adopted in these models is conceptually similar
to the extra dimensions of string theories, in the sense that one replaces
a 4-dimensional spacetime manifold with a product (or fibration), where
the fibers are the ``extra dimensions". However, unlike in string theory,
these extra dimensions are not manifolds but noncommutative spaces.
Moreover, on this product geometry encompassing spacetime directions
and noncommutative extra dimensions, one has an action functional,
the {\em spectral action}, which is a natural action functional for pure
gravity in noncommutative geometry. The key idea is that pure gravity
on a noncommutative space that is the product of a spacetime manifold
and a suitable noncommutative fiber looks like gravity coupled
to matter from the spacetime point of view. More precisely, the asymptotic
expansion for the spectral action functional delivers terms that contain
gravity terms such as the Einstein--Hilbert action and a Lagrangian for
the matter content of the model, which depends on the choice of the
finite geometry. For a simple choice of a finite dimensional algebra as
the fiber noncommutative space one recovers the Standard Model
Lagrangian \cite{CoSM}, extensions of the Minimal Standard Model that
include right handed neutrinos \cite{CCM} and supersymmetric 
extensions \cite{BroSuij}. 

\medskip

Unlike other high-energy physics models involving noncommutativity, such
as those arising in some string theory compactifications, there is here no
noncommutativity in the spacetime coordinates, only in the extra dimensions.
Moreover, these particle physics models live naturally at unification
energy. While there is, at this point, no clear picture of how the models 
should be extended to higher energies,  it has been frequently proposed that,
when moving towards the Planck scale, the model should incorporate more 
noncommutativity, which will eventually involve the spacetime coordinates as well.
Moreover, one expects that more ``seriously noncommutative" spaces (that is,
not Morita equivalent to commutative ones) should appear as one approaches
the Planck scale. In this paper we do not attempt to answer the question of
how to extend the noncommutative geometry models of gravity coupled
to matter towards the Planck scale, but we describe a toy model case for what
a behavior of the type suggested would look like, in a geometry that is
at the same time sufficiently simple to be explicitly computable, but
sufficiently complex to exhibit a nontrivial behavior.

\smallskip

Our model is constructed by adapting a very well known 
example of classical cosmologies that exhibit a chaotic 
behavior, namely the mixmaster cosmological models 
of \cite{BKL}, \cite{KLKSS}, based on the Kasner metrics
and on a discrete dynamical system related to the 
continuous fraction expansion, that governs the succession
of mixmaster cycles and Kasner epochs. We input the noncommutativity
in this model by replacing the spatial sections of this cosmology by
noncommutative tori.

\smallskip

In \S \ref{mixSec} we recall the classical mixmaster universe model,
which we formulate in the case where the spatial sections are 3-dimensional
tori $T^3$. We recall the main properties of the discrete dynamical 
system that models the mixmaster dynamics, and its relation to the
Kasner metrics and the Kasner epochs and cycles of this anisotropic
and chaotic universe model. 

\smallskip

In \S \ref{NCtoriSec} we recall some basic properties of noncommutative
3-tori, as noncommutative algebras and as spectral triples (noncommutative
Riemannian manifolds). We describe two different possible ways to
extend the mixmaster dynamical system to act on the moduli of the
noncommutative tori and not only on their metric structure. We show that
one of these models leads naturally to the occurrence of exotic smooth structures
in this noncommutative cosmological model.

\smallskip

In \S \ref{SpActSec} we discuss inflation scenarios derived from the
spectral action functional and we construct a toy model of 
deformations of parameters that gives rise to a transition from
an early universe noncommutative mixmaster cosmology to
a commutative inflationary cosmology.

\smallskip

Finally, in \S \ref{OtherSec} we discuss other aspects of the
model, related to properties of the spectral action, and we
identify physical settings in which this type of model may be
relevant.

\section{The mixmaster universe}\label{mixSec}

We review briefly in this section the basic properties of the
mixmaster universe models in general relativity, and some
aspects of the geometry that we need for the noncommutative
generalization that follows.

\smallskip
\subsection{Kasner metrics}

The mixmaster universe models were developed (see \cite{BKL}, \cite{KLKSS})
as interesting cosmological models exhibiting strong anisotropy and chaotic
behavior. One considers anisotropic metrics
\begin{equation}\label{kasnerabc}
ds^2 = - dt^2 + a(t)^2\, dx^2 + b(t)^2\, dy^2 + c(t)^2 dz^2,
\end{equation}
with the functions $a(t)$, $b(t)$, $c(t)$  of the Kasner form
\begin{equation}\label{kasner}
ds^2 = - dt^2 + t^{2p_1} dx^2 + t^{2p_2} dy^2 + t^{2p_3} dz^2,
\end{equation}
where the exponents $p_1,p_2,p_3$ satisfy the constraints
$\sum_i p_i =1$ and $\sum_i p_i^2=1$. 

\smallskip
\subsection{Discrete dynamical system}

The mixmaster universe is a solution of the Einstein equation, built
out of approximate solutions that look like Kasner metrics for certain
intervals of time (Kasner eras), with a discrete dynamical system
governing the transition from one era to the next in terms of the
change of Kasner exponents in the metric, see \cite{Mayer} and
\cite{ManMar}, \cite{Mar}.
One sets
\begin{equation}\label{piu}
\begin{array}{rl}
p_1 = & \displaystyle{\frac{-u}{1+u+u^2}} \\[4mm]
p_2 = & \displaystyle{\frac{1+u}{1+u+u^2}} \\[4mm]
p_3 = & \displaystyle{\frac{u(1+u)}{1+u+u^2}}
\end{array}
\end{equation}
The dynamics is discretized by starting, at the beginning of
each Kasner era, with a value $u_n >1$ of the parameter $u$.
Each era is then subdivided into shorter cycles, determined
by decreasing values $u_n$, $u_n-1$, $u_n -2,\ldots$. 
Within each of these cycles the metric is approximated with
a Kasner metric \eqref{kasner} with exponents \eqref{piu} 
with fixed $u=u_n-k$.  This sequence of Kasner cycles  stops
when the next value $u_n-k$ becomes smaller than one
(but still positive). Then one passes to the next Kasner era
with the transformation $u \mapsto 1/u$ and restarts the sequence
of Kasner cycles from this new value.  Thus, the transformation of
the parameter $u$ that marks the passage from the beginning of
one Kasner era to the the beginning of the next is the well known
dynamical system
\begin{equation}\label{Tun}
 T: u_n \mapsto u_{n+1} = \frac{1}{u_n - [u_n]} ,
\end{equation} 
which is the shift of the continued fraction expansion $Tx = 1/x -[1/x]$
with $x_{n+1}=Tx_n$ and $u_n =1/x_n$. 

\smallskip

Moreover, at the start of each new Kasner era, a permutation of the
three spatial directions occurs, which reassigns the role of the direction
responsible for the main dilation or contraction and of the two oscillating
directions. As shown in \cite{ManMar}, \cite{Mar}, in terms of the
dynamical system \eqref{Tun} and the shift of the continued fraction
expansion, this permutation can be described in the following way.
Identify the three spatial directions with the three points of $\bP^1(\bF_2)$
via $1\mapsto x$; $\infty \mapsto y$; $0 \mapsto z$. Then the shift
$Tx = 1/x -[1/x]$ of the continued fraction expansion on $[0,1]$ extends
to a discrete dynamical system on $[0,1]\times \bP^1(\bF_2)$ by
\begin{equation}\label{TP1F2}
T: (x,s) = \left( \frac{1}{x} - \left[ \frac{1}{x} \right], \left( \begin{array}{cc} 
-[1/x] & 1 \\ 1 & 0 \end{array}\right) \cdot s \right).
\end{equation}
This dynamical system (and its invertible two-sided extension) is
in fact the dynamical system that gives the coding of geodesics on
the modular curve $X_0(2)=\H/\Gamma_0(2)$, as shown in
\cite{ManMar}, \cite{Mar}, which leads to the description of solutions
of the mixmaster universe dynamics in terms of geodesics on this
modular curve, as recalled in Section \ref{geodsec} below.

\smallskip

Clearly, the description of mixmaster universe cosmologies in terms of this
discrete dynamical system only leads to the construction of an approximate 
solution of the Einstein equation, and one can then argue with more subtle
analytic methods in what sense this approximate solution is close to an actual solution,
but we will not be dealing with the approximation problem in this paper.

\medskip
\subsection{Mixmaster data and geodesics on the modular curve $X_0(2)$}\label{geodsec}

It was shown in \cite{ManMar}, \cite{Mar}, that the solutions of the
discretized mixmaster dynamics are in one-to-one correspondence
with geodesics on the modular curve $X_0(2)=\H/\Gamma_0(2)$,
with $\Gamma_0(2) \subset \SL_2(\Z)$ the congruence subgroup 
of level two. 

\smallskip

Recall from  \cite{ManMar}, \cite{Mar} that
every infinite geodesic on $X_0(2)$ not ending at the cusp can be coded
by data $(\omega,s)=(\omega^-,\omega^+,s)$, with $\omega^\pm \in \P^1(\R)$
and $s\in \P^1(\F_2)$, and where $\omega^\pm$ can be chosen with
$\omega^+\in [0,1]\cap (\R\smallsetminus \Q)$ and 
$\omega^-\in (-\infty,-1]\cap (\R\smallsetminus \Q)$. These can be written in terms
of continued fraction expansion as
$\omega^+=[k_0,k_1,k_2,\ldots,k_n,\ldots]$ and $\omega^-=[k_{-1};k_{-2},\ldots, k_{-n},\ldots]$
and are acted upon by the shift as
$$ T(\omega^+,s)=\left( \frac{1}{\omega^+}-\left[  \frac{1}{\omega^+} \right], 
\left( \begin{array}{cc} -[1/\omega^+] & 1 \\ 1 & 0 \end{array}\right) \cdot s \right) $$
$$ T(\omega^-,s)=\left( \frac{1}{\omega^- +[1/\omega^+]}, 
\left( \begin{array}{cc} -[1/\omega^+] & 1 \\ 1 & 0 \end{array}\right) \cdot s \right). $$
Geodesics on $X_0(2)$ are parameterized by the orbits of the data $(\omega,s)$
under the action of the shift $T$.

\smallskip

The data $(\omega,s)$ in turn determine a solution of the mixmaster dynamics,
by assigning the $[u_n]$ to be the digits $k_n$ of the continued fraction expansion
of $\omega^\pm$ and the alternation of the spatial directions being determined
by the action of $T$ on the element $s\in \P^1(\F_2)$, according to the identification
mentioned above between points of $\P^1(\F_2)$ and spatial axes. 

\smallskip
\subsection{Mixmaster tori}

Observe that the metric \eqref{kasner} can be considered equally on  a
spacetime whose spatial sections are topologically a flat space $\R^3$
or whose sections that are topologically tori $T^3$. We will focus on the
latter possibility for our noncommutative model. Thus, we assume that
the spacetime manifold is topologically a cylinder $T^3\times \R$,
endowed with the Lorentzian metric of the Kasner form \eqref{kasner}.

\smallskip

In the case of a torus $T^3$, the evolution
in time $(T^3_t, g_t)$ with the Kasner metric 
$g_t=t^{2p_1} dx^2 + t^{2p_2} dy^2 + t^{2p_3} dz^2$ has 
volume ${\rm Vol}(T^3_t)=t {\rm Vol}(T^3)$, since ${\rm Vol}(T^3_t)=t^{p_1+p_2+p_3}
{\rm Vol}(T^3)$, where we are assuming that $p_1+p_2+p_3=1$.

\smallskip

Consider the Dirac operator $\dirac_t$ on $(T^3_t, g_t)$, associated to a choice
of a spin structure $\fs$ on $T^3$. On the 3-torus $T^3$ there are eight different spin
structures $\fs_j$. We recall the following result (see \cite{Baer}, \cite{Fried}) on the
Dirac spectrum. 

\smallskip

Let $T^3=\R^3/\Lambda$ be a 3-dimensional torus, with $\Lambda$
a lattice in $\R^3$. Let $\{ \tau_1, \tau_2 , \tau_3 \}$ be a basis for $\Lambda$ and let
$\Lambda^\vee$ be the dual lattice with dual basis $\{ w_1, w_2, w_3 \}$. 
The eight spin structures are classifies by eight vectors 
$\{ \fs=(\fs_1,\fs_2,\fs_3)\,|\, \fs_i \in \{0,1\} \}$, where the value of each 
$\fs_i$ distinguishes whether
the spin structure on each of the three directions $v_i$ is twisted or untwisted.
In fact, on the circle $S^1$, the spinors for the two possible spin
structures can be identified with
$$ \Gamma(S^1,\bS)= \{ \psi: \R \to \C \,|\, \psi(t+2\pi)=\pm \psi(t) \},  $$
and the Dirac operator $-i \frac{d}{dt}$ has eigenfunctions $\psi_k(t) = \exp(2k\pi i t)$
for the trivial spin structure and $\psi_k(t)=\exp((2k+1)\pi i t)$ for the other one.

\smallskip

On $T^3=\R^3/\Z^3$, the Dirac operator $\dirac$ is of the form
\begin{equation}\label{DiracT3}
\dirac = -i (\sigma_1 \frac{\partial}{\partial x} + \sigma_2 \frac{\partial}{\partial y} + \sigma_3
\frac{\partial}{\partial z}) = \left( \begin{array}{cc}  -i \frac{\partial}{\partial z} & 
- \frac{\partial}{\partial y}  -i  \frac{\partial}{\partial x}  \\ \frac{\partial}{\partial y}  
-i  \frac{\partial}{\partial x} &  i \frac{\partial}{\partial z} \end{array} \right).
\end{equation}
More generally, on $T^3=\R^3/\Lambda$, one can write the Dirac operator in
the form (up to a possible overall additive shift)
\begin{equation}\label{DiracT3tau}
\dirac =-i \sum_{j=1}^3 (\tau_j \cdot \underline{\partial} ) \sigma_j = -i((\tau_1 \cdot \underline{\partial} ) \sigma_1 + (\tau_2 \cdot \underline{\partial} ) \sigma_2 + (\tau_3 \cdot \underline{\partial} ) \sigma_3 )
\end{equation}
$$ = \left( \begin{array}{cc}  -i \partial_{\tau_3} & 
- \partial_{\tau_2}  -i  \partial_{\tau_1} \\  \partial_{\tau_2} 
-i  \partial_{\tau_1} &  i \partial_{\tau_3} \end{array} \right), $$
where $\underline{\partial}=(\partial_1,\partial_2,\partial_3)=(\frac{\partial}{\partial x},
\frac{\partial}{\partial y},\frac{\partial}{\partial z})$, 
$\partial_{\tau_j}=\tau_j\cdot \underline{\partial}$, and $\sigma_j$ are the Pauli
matrices.

\smallskip

Then on the 3-torus $T^3=\R^3/\Lambda$, 
the Dirac operator $\dirac$ on the spin structure $\fs$ has spectrum
\begin{equation}\label{specdiracj}
\Spec(\dirac) =\{ \pm 2\pi \| w + \frac{1}{2} \sum_{j=1}^3 \fs_j w_j \| \,|\, 
w\in \Lambda^\vee\}.
\end{equation}

In particular, in the case of the torus $T^3_t=\R^3/\Lambda_t$, 
where the dual lattice $\Lambda_t^\vee$ is spanned by the
basis of vectors $\{ t^{p_1} e_1, t^{p_2} e_2, t^{p_3} e_3 \}$, with $e_i$ the standard
orthonormal basis, the Dirac operator is given by 
\begin{equation}\label{DiracT3t}
\dirac = -i (\sigma_1 t^{-p_1} \frac{\partial}{\partial x} + \sigma_2 t^{-p_2}
\frac{\partial}{\partial y} + \sigma_3 t^{-p_3}
\frac{\partial}{\partial z}) 
\end{equation}
$$ = \left( \begin{array}{cc}  -i  t^{-p_3} \frac{\partial}{\partial z} & 
- t^{-p_2}\frac{\partial}{\partial y}  -i  t^{-p_1} \frac{\partial}{\partial x}  \\ 
t^{-p_2} \frac{\partial}{\partial y}  -i  t^{-p_1} \frac{\partial}{\partial x} 
&  i  t^{-p_3} \frac{\partial}{\partial z} \end{array} \right) . $$

\smallskip

The spectrum of $\dirac_t$, for the spin structure $\fs$, will then be of the form
\begin{equation}\label{specdiracjt}
\Spec(\dirac_t) =\{ \pm 2\pi \| (t^{-p_1} k, t^{-p_2} m, t^{-p_3} n) 
+ \frac{1}{2} \sum_{j=1}^3 \fs_j t^{-p_j}e_j \| \,|\, 
(k,m,n)\in \Z^3 \}.
\end{equation}

\smallskip
\subsection{Mixmaster dynamics on classical tori}

The dependence of the Kasner exponents on the $u$-parameter
as in \eqref{piu} and the discrete dynamical system \eqref{Tun},
with the permutations \eqref{TP1F2}, 
correspondingly determine a sequence of 3-tori $T^3_{u_n}$
with Dirac operators as in \eqref{specdiracjt}, where the
exponents $p_i$ are functions of $u=u_n$ through \eqref{piu}
and the permutation \eqref{TP1F2} of the coordinate axes.

\smallskip

As in Section \ref{geodsec}, passing of each Kasner era is marked by the transition
$u_n \mapsto u_{n+1}$ and $s_n \mapsto s_{n+1}$, where
$u_n =1/x_n$ and $x_{n+1}=Tx_n$, and with $s_n \in \P^1(\F_2)$,
with $s_{n+1} = Ts_n$ with the map $T$ as in  \eqref{TP1F2}. 
In particular, the action of $T$ that gives the transition from the
$n$th to the $(n+1)$st Kasner era is given by the action of the
matrix
\begin{equation}\label{GL2Z}
\gamma_n= \left(\begin{array}{cc} -[u_n] & 1 \\ 1 & 0  \end{array}\right) \in \GL_2(\Z),
\end{equation} 
which simultaneously acts on changing the metric of the torus and
on permuting its three generators.

\bigskip

\section{Noncommutative 3-tori and mixmaster cosmologies}\label{NCtoriSec}

We now reconsider the model of mixmaster universe described
above, in a setting where the three spatial coordinates that give
the 3-torus $T^3_t$ are replaced by a {\em noncommutative} 3-torus
$\bT^3_\Theta$. We describe these noncommutative spaces as
Riemannian geometries, in the noncommutative sense, that is,
as {\em spectral triples}. We then propose different possible ways
in which the discrete dynamical system that describes the
evolution of a classical mixmaster cosmology can be extended
to involve also an action on the parameters (the moduli) of the
noncommutative torus itself, so that not only the metric structure,
but also the underlying noncommutative space, evolves along
with the succession of Kasner epochs.

\subsection{3-tori as spectral geometries}

In noncommutative geometry the analog of a Riemannian spin manifold
is described by the data of a {\em spectral triple}. These consist of 
a triple $(\cA,\cH,D)$, where $\cA$ is an involutive, dense
subalgebra of a $C^*$-algebra closed under holomorphic functional
calculus, together with a (faithful) representation $\pi: \cA \to \cB(\cH)$ as
bounded operators on a separable Hilbert space $\cH$, and a ``Dirac
operator" $D$. The latter is a self-adjoint (unbounded) operator, densely
defined on $\cH$, with compact resolvent and satisfying the
condition of having bounded commutators with
the elements of the algebra, $[\pi(a),D] \in \cB(\cH)$, for all $a\in \cA$.

\smallskip

A smooth compact Riemannian spin manifold $X$ is a special case of
a spectral triple, where the data $(\cA,\cH,D)$ are given by $\cA=\cC^\infty(X)$,
$\cH=L^2(X,\bS)$, with $\bS$ the spinor bundle, and $D=\dirac_X$ the
Dirac operator. 

\smallskip

Thus, the mixmaster tori $T^3_t=\R^3/\Lambda_t$ described above are spectral triples
with $\cA=\cC^\infty(T^3)$, and with $\cH=L^2(T^3_t,\bS_{\fs})$, where
$\bS_{\fs}$ is the spinor bundle for the spin structure $\fs=(\fs_1,\fs_2,\fs_3)$
as above, with
\begin{equation}\label{ST3s}
L^2(T^3_t,\bS_{\fs})=\{ \psi: \R^3 \to \C^2\, |\, \psi\in L^2, \,\,\, 
\psi(\underline{x} + t^{p_j} v_j)=(-1)^{\fs_j} \psi(\underline{x}) \},
\end{equation}
for $\underline{x}=(x,y,z) \in \R^3$ and $\{ \tau_j \}$ the basis of $\Lambda$,
and with Dirac operator 
\begin{equation}\label{DiracT3taut}
\dirac =-i \sum_{j=1}^3 t^{-p_j} (\tau_j \cdot \underline{\partial} ) \sigma_j =-i( t^{-p_1} (\tau_1 \cdot \underline{\partial} ) \sigma_1 + t^{-p_2} (\tau_2 \cdot \underline{\partial} ) \sigma_2 + t^{-p_3}(\tau_3 \cdot \underline{\partial} ) \sigma_3 ) .
\end{equation}
Here we scaled the spatial coordinates $x_i \mapsto t^{p_i} x_i$, so that
$\partial_{x_i}\mapsto t^{-p_i} \partial_{x_i}$, and fixed the $\tau_i$, instead of
scaling $\Lambda \mapsto \Lambda_t$ as before:  the resulting
\eqref{DiracT3taut} is the same.

\medskip
\subsection{Noncommutative 3-tori and their spectral geometry}

A noncommutative 3-torus is the universal $C^*$-algebra $\cA_\Theta$ generated by three
unitaries $U_1$, $U_2$, $U_3$ with the relations
\begin{equation}\label{UjUk}
U_j U_k = \exp(2\pi i \Theta_{jk}) U_k U_j,
\end{equation}
where $\Theta=(\Theta_{jk})$ is a skew-symmetric matrix
\begin{equation}\label{Theta}
\Theta =\left( \begin{array}{ccc}  0 & \theta_3 & -\theta_2 \\
-\theta_3 & 0 & \theta_1 \\
\theta_2 & -\theta_1 & 0 
\end{array}\right) .
\end{equation}
It can also be described as the twisted group $C^*$-algebra $C^*_r(\Z^3,\sigma_\Theta)$,
where the $U(1)$-valued 2-cocycle $\sigma_\Theta$ is given by (see \cite{Elliott})
\begin{equation}\label{sigmaTheta}
\sigma_\Theta(\underline{x},\underline{y}) = \exp(\pi i \langle \underline{x}, \Theta \underline{y}\rangle).
\end{equation}
A detailed discussion of the main properties of noncommutative 3-tori
can be found in \cite{Bedos}.

\medskip

Recall that, for a noncommutative 3-torus, the algebra of
smooth functions is defined as 
\begin{equation}\label{AinftyTheta}
\cA^\infty_\Theta=\{ X\in \cA_\Theta, \, X=\sum_{m,n,k \in \Z} a_{m,n,k} \, U_1^m U_2^n U_3^k \,|\, 
a=(a_{m,n,k}) \in \cS(\Z^3,\C) \},
\end{equation}
which is the noncommutative analog of functions on the 3-torus with
rapidly decaying Fourier coefficients.

\medskip

It was recently proved by Venselaar in \cite{Ven} that all equivariant real spectral
triples on rank $n$ noncommutative tori are isospectral deformations,
in the sense of \cite{CoLa}, where spin structures are Dirac operators on commutative
flat $n$-dimensional tori $T^n$. The case of rank-two tori was previously shown
in \cite{PaSi}. 

\smallskip

Thus, these spectral triples will all be of the form $(\cA_\Theta^\infty, L^2(T^3,\bS), \dirac)$,
where $\bS$ is the spinor bundle for one of the eight spin structures on the ordinary torus $T^3$
and the Dirac operator $\dirac$ is of the form \eqref{DiracT3tau} (up to an overall additive
constant) with $\{ \tau_j \}_{j=1,2,3}$ a basis.

\subsection{Noncommutative 3-tori, moduli, and mixmaster evolution}

Now we consider again the discrete dynamical system of mixmaster evolution,
where the passage from one Kasner era to the next is determined by the action
of the matrix  $\gamma_n \in \GL_2(\Z)$ of \eqref{GL2Z}. In this noncommutative
setting, in addition to the action on the exponents of the Kasner metric and
the permutation of the spatial directions (here given by a permutation of the
three generators of the noncommutative torus algebra) and by the corresponding
action on the Dirac operator in \eqref{DiracT3taut} as in the commutative case, one also 
has the modulus $\Theta$ of the
noncommutative torus, which is trivial in the commutative case. Thus, one
can propose extensions of the mixmaster dynamics where the modulus
$\Theta$ is also acted upon in the transition from one Kasner epoch to
the other. We present in the following subsections examples of two possible
such extensions of the mixmaster dynamics and we illustrate some of
their properties. As we discuss below, these will have some interesting 
consequences on the properties of the noncommutative mixmaster 
cosmologies that differ from their classical counterparts.

\medskip
\subsection{Moduli evolution by auxiliary mixmaster data}

In this scenario, we assume given a choice of mixmaster data, by which
we mean a choice of a geodesic on the modular curve $X_0(2)$, or
equivalently a choice of data $(\omega,s)$ up to the action of the shift $T$
as recalled in Section \ref{geodsec}. This means that we have an assigned
sequence of matrices $\gamma_n \in \GL_2(\Z)$, of the form \eqref{GL2Z}.
We also have a sequence $s_n$ of elements in $\P^1(\F_2)$, which determines,
in each Kasner era, which spatial direction is the dominant direction driving
expansion or contraction.

\smallskip

Recall then that, if we write the modulus for the noncommutative $3$-torus
as the vector $\underline{\theta}=(\theta_1,\theta_2,\theta_3)\in \R^3$ 
of the three parameters out of which the skew-symmetric matrix $\Theta$
of \eqref{Theta} is built, then for any matrix $M \in \GL_3(\Z)$ we have an action
$\underline{\theta}'=M\underline{\theta}$ and we let $\Theta'$ be the
skew-symmetric matrix corresponding to the new values $\underline{\theta}'$.

\smallskip

The fact that, in each era of the era of the mixmaster dynamics, there is one of the
three spatial directions that dominates the contraction or expansion, identified by
the given sequence $s_n$ allows us then to define
an action of the matrices $\gamma_n \in \GL_2(\Z)$ on the parameters of
the noncommutative torus,  by embedding $\GL_2(\Z)$ inside $\GL_3(\Z)$,
so that it acts as the identity on the parameter associated to the dominant 
direction and as $\gamma_n$ on the other two. Since this action is accompanied
by a permutation of the directions, at the following change of Kasner era the
dominant direction will have changed and the embedding of 
$\GL_2(\Z)$ inside $\GL_3(\Z)$ used in defining the action on $\underline{\theta}$
will change accordingly.

\smallskip

Thus, for example, if at the $n$th Kasner era the first coordinate is the 
dominant direction, we obtain the transformation
\begin{equation}\label{thetatransf}
 \underline{\theta}=(\theta_1,\theta_2,\theta_3) \mapsto 
 \underline{\theta}'=(\theta_1, -[u_n]\theta_2 +\theta_3, \theta_2). 
\end{equation} 

\smallskip

As we discuss in the next subsection, the choice of this model
for mixmaster evolution of the noncommutative torus moduli
has an interesting consequence: along the sequence of Kasner
epochs, not only the metric structure of the noncommutative tori
undergoes a sequence of transformations analogous to the
classical mixmaster dynamics, but at the same time the {\em smooth
structure} of the noncommutative $3$-torus undergoes a sequence
of transitions to different {\em exotic structures}.

\subsection{Smooth structures and Kasner eras}

In ordinary commutative geometry, the first occurrence of
exotic smooth structures, meaning examples of 
smooth manifolds that are homeomorphic but not diffeomorphic,
occurs in dimension four. In noncommutative geometry, however,
the simplest example of exotic smooth structures is known to 
occur already in three dimensions, for noncommutative 3-tori.

\smallskip

We consider here the algebra $\cA^\infty_\Theta$ of smooth functions
described in \eqref{AinftyTheta}.

\smallskip

One can then introduce the following two equivalence relations on the
modulus $\Theta$ (see \cite{Bedos}):
\begin{itemize}
\item $\Theta \sim \Theta'$ $\Leftrightarrow$ $\underline{\theta}'=M \, \underline{\theta}$, with $M\in \SL_3(\Z)$;
\item  $\Theta \approx \Theta'$ $\Leftrightarrow$ $\underline{\theta}'=M \, \underline{\theta}$, with $M\in \GL_3(\Z)$.
\end{itemize}
Then (see \cite{Bedos}) one has algebra isomorphisms:
\begin{equation}\label{isomATheta}
\cA_\Theta \simeq \cA_{\Theta'} \ \ \Leftrightarrow \ \  \Theta\approx \Theta' \ \ \ \text{ while } \ \ \
\cA^\infty_\Theta \simeq \cA^\infty_{\Theta'} \ \ \Leftrightarrow \ \  \Theta\sim \Theta'.
\end{equation}
The $C^*$-algebra case follows from \cite{Brenken}, \cite{Bedos} and \cite{EllLin}, while
the smooth subalgebras case uses \cite{Brenken}, \cite{BCEN}, \cite{CEGJ}.

\smallskip

One sees from this result that the noncommutative 3-tori have exotic smooth structures:
any two tori with $\Theta\approx \Theta'$ through a matrix $\underline{\theta'}=M \underline{\theta}$ with $M\in \GL_3(\Z)$ but $M\notin \SL_3(\Z)$ are homeomorphic (in the sense that the
algebras of continuous functions are isomorphic) but not diffeomorphic (in the sense that
the algebras of smooth functions are not isomorphic).

\smallskip

Thus, an interesting phenomenon happens in the mixmaster dynamics, whereby the
passage to each successive Kasner era, which is determined by the action of a matrix
$M\in \GL_3(\Z)$, which has $\det(M)=\det(\gamma_n)=-1$, changes the torus $\bT^3_\Theta$
to a new torus $\bT^3_{\Theta'}$ which is homeomorphic, but with a different smooth
structure.  

\smallskip

The topic of exotic smooth structures and their relevance to physics
has been explored in various aspects in recent years, see for
instance \cite{AssMa}, \cite{Dust}. This simple observation about
the noncommutative tori shows that, when allowing noncommutativity
in the space coordinates, one can more easily encounter phenomena 
involving exotic smoothness, such as, in this case, changes of smooth 
structure.

\smallskip

\subsection{Moduli evolution by internal mixmaster data}

We propose here another possible way to extend to the
torus moduli the discrete dynamical system defining the
mixmaster dynamics. In this case, one does not assume
a given classical mixmaster solution, but constructs it
directly in terms of the torus moduli themselves.

\smallskip

In this case, to define the action of the mixmaster dynamics
on the modulus $\Theta$, we
recall the following equivalent description of the noncommutative
3-torus $\cA_\Theta$, see \cite{Bedos}. One can view the 3-dimensional
noncommutative torus $\cA_\Theta$ as a crossed product $C^*$-algebra
for an action of $\Z$ on a 2-dimensional noncommutative torus:
\begin{equation}\label{3toruscross}
\cA_\Theta = \cA_{\theta_3}\rtimes_\alpha \Z,
\end{equation}
where $\cA_{\theta_3}$ is the 2-dimensional noncommutative
torus generated by two unitaries $U$ and $V$ with the relation
$VU = e^{2\pi i \theta_3} UV$, and the action $\alpha:\Z \to {\rm Aut}(\cA_{\theta_3})$
is given by
$$ \alpha(U) = e^{2\pi i \theta_2} U,  \ \ \  \alpha(V) = e^{-2\pi i \theta_1} V . $$

\smallskip

One can then change the parameters of the noncommutative 3-torus, in 
passing to the next Kasner era, by acting on the 2-dimensional noncommutative
2-torus $\cA_{\theta_3}$ by a Morita equivalence, implementing the change
of parameter given by the action of the matrix $\gamma_n$ on $\theta_3$ by
fractional linear transformations
\begin{equation}\label{GL2Zact2}
 \theta_3 \mapsto \gamma_n(\theta_3)=\frac{ -[u_n]\theta_3 +1}{\theta_3} =\frac{1}{\theta_3} -[u_n], 
\end{equation} 
for $\gamma_n$ as in \eqref{GL2Z}, but where now the integers $k_n=[u_n]$ are the
digits of the continued fraction expansion of $\theta_3$ itself.

\medskip

In the particular case where the parameter $\theta_3$ is a quadratic irrationality,
namely an irrational number that is contained in some real quadratic field embedded
in $\R$, then the digits of the continued fraction expansion of $\theta_3$ are
eventually periodic, and there is a natural choice of the mixmaster data $(\omega,s)$,
with $\omega^\pm=\{ \theta_3, \theta_3' \}$, with $\theta_3'$ the Galois conjugate of
$\theta_3$. This choice corresponds to a closed geodesic in $X_0(2)$.

\medskip

Notice that here we are acting by Morita equivalences of the 2-dimensional
noncommutative torus $\cA_{\theta_3}$, which are implemented by an action of $\GL_2(\Z)$,
while for 3-dimensional smooth noncommutative tori in the generic case, the 
Morita equivalences are implemented  by the action of the group
$SO(3,3|\Z)$, see \cite{EllLi1}, \cite{RiefSchw} and \cite{EllLi2} for a 
complete classification up to Morita equivalences.

\bigskip

\section{The spectral action and inflation scenarios}\label{SpActSec}

We now consider the action functional, the {\em spectral action}, for the
noncommutative mixmaster cosmologies.

The spectral action functional is defined as $\Tr (f(D/\Lambda))$,
where $\Lambda$ is the energy scale and $f$ is a test function, usually a 
smooth approximation of a cutoff function. 
This is regarded as a spectral formulation of gravity in noncommutative
geometry.  This action functional has an asymptotic expansion at high
energies (see \eqref{spactAexpand} below). Thus, one can approach the
computation of this action functional either by explicit information on the
spectrum and a computation of the series defined by $\Tr (f(D/\Lambda))$
(as in \cite{ChCo}, \cite{ChCo2}, \cite{CMT}, \cite{MPT1}, \cite{MPT2}), 
or else through its asymptotic expansion and the computation via heat kernel
methods and local expressions in curvature tensors of the various terms 
in the expansion, as, for instance, in \cite{CMT}, \cite{CCM}, \cite{EILS},
\cite{GIV}.

\medskip

The computation of the terms in the asymptotic expansion shows (see
for instance \cite{CCM}) that one recovers the terms in the usual
classical action for gravity, namely the Einstein--Hilbert action and the
cosmological term, together with other ``modified gravity" terms, such
as a Weyl curvature conformal gravity term. In addition, depending on
the possible introduction of a fiber over spacetime given by a finite
noncommutative geometry, one obtains further terms that give a
Larangian for matter minimally coupled to gravity. This can be the
Lagrangian of the minimal standard model (see \cite{CoSM}) or of the
standard model with additional right handed neutrinos with 
Majorana mass terms (see \cite{CCM}) or models with supersymmetry
(see \cite{BroSuij}).  In addition to the minimal coupling of matter to
gravity one also finds terms such as a non-minimal conformal
coupling of the curvature to the Higgs field.

\medskip

All this shows that one can use the spectral action as an action
functional either for pure gravity (on a commutative manifold) or
for gravity coupled to matter on a product geometry. The main
philosophy behind it is that pure gravity on a noncommutative
space can manifest itself as gravity coupled to matter when seen
from a commutative point of view. 

\medskip

The asymptotic expansion of the spectral action was also computed 
explicitly for truly noncommutative spaces like noncommutative tori, see
\cite{EILS}, \cite{GIV}.

\medskip

We work here under the assumption that the spectral action is
our modified gravity model, for commutative and noncommutative 
geometries alike and we discuss its behavior in a mixmaster
case where the underlying spatial slices are noncommutative 3-tori.

\medskip

We also discuss possible slow-roll inflation scenarios derived
from the spectral action and how they affect and interfere with
the underlying mixmaster dynamics.

\medskip
\subsection{Slow-roll inflation in anisotropic cosmologies}

In the usual isotropic Friedmann cosmologies with Lorentzian metrics
of the form
$$ ds^2 = -dt^2 + a(t)^2 (dx^2 + dy^2 +dz^2), $$
inflation is an accelerated expansion of the universe that
corresponds to the scale factor $a(t)$ satisfying $\ddot{a}>0$.
The Friedmann equation relates the scale factor to the Hubble
parameter
$$ \frac{\dot{a}}{a} = H, $$
and the slow-roll models of inflation are based on the relation
of the latter to a scalar field $\phi$ with potential $V(\phi)$ via
$$ H^2 = \frac{8 \pi G}{3} (\frac{1}{2} \dot{\phi}^2 +V(\phi)). $$
The slow-roll condition then corresponds to the condition that $\dot{\phi}^2 << |V(\phi)|$,
so that the term in $\dot{\phi}$ in the Friedmann equation becomes negligible. 

\medskip

In the case of anisotropic spacetimes of the form \eqref{kasnerabc}, 
one introduces an average scale factor
\begin{equation}\label{avscale}
 \fa(t) = (a(t) b(t) c(t))^{1/3} 
\end{equation}
and directional Hubble parameters
$$ H_1 = \frac{\dot{a}}{a}, \ \ \  H_2 =\frac{\dot{b}}{b}, \ \ \  H_3 =\frac{\dot{c}}{c}, $$
and an average Hubble parameter
$$ H = \frac{1}{3} (H_1 + H_2 + H_3) . $$
This satisfies
$$ H = \frac{1}{3} ( \frac{\dot{a}}{a} +\frac{\dot{b}}{b} +\frac{\dot{c}}{c} ) = 
\frac{1}{3} \frac{(\dot{a} bc + \dot{b} ac + \dot{c} ab)}{(abc)^{2/3}}\cdot \frac{1}{(abc)^{1/3}} =
 \frac{\dot{\fa}}{\fa}. $$
Thus, we obtain the same picture as in the isotropic case, but for the average Hubble
parameter and the average scale factor.

\medskip

In the case of a mixmaster cosmology \eqref{kasner}, where $p_1+p_2+p_3=1$,
the average scale factor is just given by $\fa =(t^{p_1+p_2+p_3})^{1/3} =t^{1/3}$, with
$\dot{\fa}/\fa=(1/3) t^{-1}$.  Thus, the Friedmann equation for the Hubble parameter in 
a pure mixmaster dynamics is of the form
\begin{equation}\label{Hubblemix}
  \frac{\dot{\fa}}{\fa}  =H=  \frac{1}{3} t^{-1},
\end{equation}
or equivalently $H_1+H_2+H_3=t^{-1}$.

\smallskip

For a mixmaster cosmology, one also has
$$ \frac{\ddot{a}}{a} + \frac{\ddot{b}}{b} + \frac{\ddot{c}}{c} =0, $$
since, for $a(t)=t^{p_1}$, one has $\ddot{a}/a=p_1(p_1-1) t^{-2}$, and similarly
for the scale factors $b(t)$ and $c(t)$ so that
$$ \frac{\ddot{a}}{a} + \frac{\ddot{b}}{b} + \frac{\ddot{c}}{c} = (p_1(p_1-1)+p_2(p_2-1)+
p_3(p_3-1)) t^{-2}, $$
which vanishes, since we assume $p_1+p_2+p_3=1$ and $p_1^2+p_2^2+p_3^2=1$.

\medskip
\subsection{The nonperturbative spectral action for 3-tori}

Since we are dealing with spectral triples that are isospectral
deformations of commutative tori, we can refer to the computation
of the nonperturbative spectral action obtained in \cite{MPT1} for
3-dimensional flat tori, using the Poisson summation formula
as in \cite{ChCo2}.

\smallskip

Proceeding as in Theorem 8.1 of \cite{MPT1}, for a torus $T^3$ with
metric $a(t)^2 dx^2+ b(t)^2 dy^2 + c(t)^2 dz^2$ and Dirac operator
\begin{equation}\label{diracabc}
\dirac_t = -i (\sigma_1 \frac{1}{a(t)} \frac{\partial}{\partial x} +
\sigma_2 \frac{1}{b(t)} \frac{\partial}{\partial y} + \sigma_3 \frac{1}{c(t)} \frac{\partial}{\partial z}),
\end{equation}
the spectral action is of the form
\begin{equation}\label{spactT3abct}
\Tr( f(\dirac_t^2/\Lambda^2)) = a(t) b(t) c(t)
\frac{\Lambda^3}{4\pi^3} \int_{\R^3} f(u^2+v^2+w^2)\, du\, dv\, dw \,
+ O(\Lambda^{-k}), 
\end{equation} 
for arbitrary $k>0$.

Thus, for the mixmaster torus $T^3_t$ with Dirac operator $\dirac_t$
as in \eqref{DiracT3t}, one finds, independently of the spin structure,
\begin{equation}\label{spactT3t}
\Tr( f(\dirac_t^2/\Lambda^2)) =t\cdot 
\frac{\Lambda^3}{4\pi^3} \int_{\R^3} f(u^2+v^2+w^2)\, du\, dv\, dw \,
+ O(\Lambda^{-k}) ,
\end{equation} 
since in this case $p_1+p_2+p_3=1$, so that $a(t) b(t) c(t)=t^{p_1+p_2+p_3}=t$.

\smallskip

Notice that the factor $a(t) b(t) c(t) \Lambda^3/(4\pi^3)$ in \eqref{spactT3abct}
behaves exactly like the isotropic case, when one introduces the average scale
factor $\fa(t) =(a(t) b(t) c(t))^{1/3}$ as discussed in the previous subsection. 
In terms of the average scale factor the spectral action has the form
\begin{equation}\label{spactT3aver}
\Tr( f(\dirac_t^2/\Lambda^2)) = 
\frac{\fa(t)^3 \Lambda^3}{4\pi^3} \int_{\R^3} f(u^2+v^2+w^2)\, du\, dv\, dw \,
+ O(\Lambda^{-k}).
\end{equation}

\medskip
\subsection{Noncommutative mixmaster cosmologies and inflation}

We can then adapt the same analysis used in \cite{MPT1}, \cite{MPT2}, \cite{CMT}
to obtain a slow-roll potential out of a perturbation of the spectral action. To that
purpose, we compute the spectral action in 4-dimensions, on a product $T^3_t\times S^1$,
with a compactified direction $S^1$ of size $\beta$ (after passing to a Wick rotation to Euclidean space and a compactification), with the product Euclidean metric.  The replacement $D^2\mapsto D^2+\phi^2$ produces a potential for the field $\phi$, which, for sufficiently small values of the parameter $x=\phi^2/\Lambda^2,$ recovers the usual shape of a quartic potential for the 
field $\phi$, conformally coupled to gravity. 
This gives, as in \cite{MPT1}, 
\begin{equation}\label{spact4d}
\Tr(h(D^2/\Lambda^2))= \frac{\Lambda^4 \beta a(t)b(t)c(t)}{4\pi} \int_0^\infty u h(u)\, du + O(\Lambda^{-k}) .
\end{equation}
The perturbation is then computed as 
\begin{equation}\label{TrhDphi}
\Tr(h((D^2+\phi^2)/\Lambda^2)) = \Tr(h(D^2/\Lambda^2)) +\frac{\Lambda^4 \beta a(t)b(t)c(t)}{4\pi}
\cV(\phi^2/\Lambda^2) ,
\end{equation}
with $\cV(x)=\int_0^\infty u (h(u+x)-h(u)) \, du$, for $x=\phi^2/\Lambda^2$ where the last term $$\frac{\Lambda^4 \beta a(t)b(t)c(t)}{4\pi} \cV(\phi^2/\Lambda^2)$$ determines the 
slow-roll potential $V(x)$. 

\smallskip

For a Kasner metric, this term is of the form 
\begin{equation}\label{cVt}
 V(x)= \frac{\Lambda^4 \beta \, \fa(t)^3}{4\pi} \cV(x), 
\end{equation} 
with $\fa(t)$ the average scale factor. This
calls for a discussion of the $t$-dependence of the
other parameters, $\Lambda$ and $\beta$, in this expression. 

\medskip
\subsection{Time dependence of the parameters}\label{timesec}

We have argued in \cite{MPT1} and \cite{MPT2} that in the case of an 
isotropic expanding cosmology with a single scale factor $a(t)$, 
the energy scale associated to the cosmological timeline
should behave like $\Lambda(t)\sim 1/a(t)$. The possible interpretations of a 
time dependence for the parameter $\beta$ are less easily justified, as this
parameter is an artifact of the $S^1$-compactification. We refer the reader to
section 3.1 of \cite{MPT2} for a discussion of the interpretation of this parameter
as an inverse temperature and its relation to the parameter $\Lambda$.

\smallskip

In view of the expression of the spectral action \eqref{spactT3aver}, in
terms of the average scale factor $\fa(t) =(a(t) b(t) c(t))^{1/3}$, it seems
natural to apply the same reasoning on the time dependence of 
the parameter $\Lambda$ and $\beta$, as in \cite{MPT2}, in terms 
of $\fa(t)$. This means, for example, that one expects the energy
scale $\Lambda$, along the cosmological timeline, to behave like
$\Lambda(t) \sim 1/\fa(t)$. 
  
\smallskip

In the usual setting of isotropic Friedmann cosmology, one relates the
scale factor and the Hubble parameter to a density function, 
\begin{equation}\label{Hubblerho}
\left( \frac{\dot{a}}{a} \right)^2  = H^2 = \frac{8\pi G}{3} \rho(t), 
\end{equation}
where the form of the density function depends on the various
cosmological eras: in a modern matter-dominated universe,
the density function behaves, as a function of time, like $\rho(t)\sim a(t)^{-3}$,
which gives an evolution of the scaling factor with power law $a(t) \sim t^{2/3}$;
in a radiation-dominated universe the density function behaves like 
$\rho(t)\sim a(t)^{-4}$, and consequently the scaling factor evolves as $a(t)\sim t^{1/2}$; 
while in a vacuum-dominated universe, as in the inflation epoch, $\rho(t)$ 
is constant and $a(t)$ is growing exponentially. In a slow-roll model of inflation, 
the constant is given by the plateau value of the slow-roll potential.

\smallskip

The latter observation implies that, in the expression \eqref{cVt} for the potential
obtained from the spectral action, assuming as above that $\Lambda$ and $\fa$
have an inverse dependence on time, the consistency with \eqref{Hubblerho}
would then suggest that the parameter $\beta$, introduced artificially
in the model as a radius of Euclidean compactification, should have a time
dependence that is determined by the behavior of the density function 
$\beta(t)\sim \rho(t)$ (up to a multiplicative constant).

\smallskip

As we will see in \S \ref{damp} below, it is convenient to consider
different possible $t$-dependences of the parameter $\beta$,
as these make it possible to model transitions between different
regimes in the very early universe (see also \cite{MaPie} for other
discussions of time dependence of parameters in the very early
universe in cosmological models based on the spectral action).

\smallskip

Notice that, in any case, the time dependence (and the dependence 
on $\beta$ and $\Lambda$) is only in the amplitude multiplicative
factor of \eqref{cVt}, and it disappears entirely when one computes 
the slow-roll parameters, which depend only on the ratios $V^\prime/V$ and
$V^{\prime\prime}/V$ (see \cite{MPT1} and \cite{MPT2}), through the
expressions
$$ \epsilon = \frac{M_{Pl}^2}{16\pi} \left( \frac{V^\prime}{V} \right)^2, \ \ \ 
\eta = \frac{M_{Pl}^2}{8\pi} \frac{V^{\prime\prime}}{V}. $$

\medskip
\subsection{The scalar field potential and the mixmaster dynamics}\label{Vmixsec}

In the presence of a slow-roll potential,
an initial mixmaster dynamics gets altered: the coupling of the potential 
to the average scale factor via the (anisotropic) Friedmann equation
disrupts the mixmaster oscillations. This phenomenon was already well
known in the case of the classical mixmaster dynamics (see for
 instance chapter 8 of \cite{PriCosm}), where it is used to argue that 
 the potential provides a mechanism of transition from
an anisotropic chaotic system to a standard big bang singularity.

\medskip

We discuss here the effect of the slow-roll potential associated to the
spectral action on the underlying mixmaster dynamics, and on the
corresponding transformation of the torus moduli.

\medskip

It is easy to see why, already in the classical case, the presence 
of a slow-roll potential disrupts a mixmaster dynamics. In fact,
suppose given, at time $t_0$, an initial condition given by the
Kasner metric of a Kasner epoch in a mixmaster universe. We have
$a(t_0)=t_0^{p_1}$, $b(t_0)=t_0^{p_2}$, $c(t_0)=t_0^{p_3}$. 
Suppose that a slow-roll potential is turned on, with value near the
plateaux level. Then the Friedmann equation predicts an evolution
of the average scale factor by
$\fa(t) = \fa(t_0) \exp(\gamma (t-t_0))$, where
$\gamma = \sqrt{(8\pi G V_\infty)/3}$, with $V_\infty$ the plateaux
value of the potential. This is compatible, for example, with
a solution for the individual scale factors of the form
\begin{equation}\label{expai}
\begin{array}{rl}
a(t)= & t_0^{1/3} t^{p_1-1/3} \exp(\gamma (t-t_0)), \\
b(t)= & t_0^{1/3} t^{p_2-1/3} \exp(\gamma (t-t_0)),  \\
c(t)= & t_0^{1/3} t^{p_3-1/3} \exp(\gamma (t-t_0)), 
\end{array}
\end{equation}
where the exponential factor dominates over the underlying mixmaster
dynamics.

\medskip
\subsection{Damping effects}\label{damp}

It would be interesting to further study if one can use the presence of a slow-roll
potential associated to the spectral action to model a transition from a noncommutative
early universe to a commutative modern universe, through a simultaneous
``override" of the mixmaster oscillations both for the metric parameters and for 
the noncommutative torus moduli.

\smallskip

In fact, according to the general philosophy about noncommutativity
in the spacetime directions, one expects it to appear only close to
the Planck scale (or the Planck era in the cosmological timeline),
while near the unification scale (and therefore during the
inflationary epoch, which is located between the unification
and the electroweak epoch), the universe is already exhibiting
noncommutativity only in the extra dimensions of the finite
geometry that describe matter and forces, but not in the 
spacetime directions. Thus, one expects to find some mechanisms
that damp the noncommutativity before the slow-roll inflation mechanism 
becomes relevant.

\medskip

For example, one can conceive of the following type of toy model scenario,
where one obtains in two steps
a transition from a mixmaster dynamics with noncommutative tori
in the very early universe to a slow-roll inflationary cosmology with
ordinary commutative tori. The first step is a deformation of noncommutative
tori to commutative tori, obtained via a deformation of the Kasner exponents, 
which disrupts the mixmaster dynamics but leaves
the Friedmann equation unaltered, and then 
a second step, in a purely commutative setting, where a deformation of
the parameter $\beta$ affects a second transition from a regime based on 
the Friedmann equation for a Kasner metric to the Friedmann equation for
an inflationary cosmology with slow-roll potential determined by the 
spectral action.

\medskip
\subsection{Deforming the Kasner parameters}\label{dampsec2}

Suppose that the dependence of the exponents $p_i(u)$ of \eqref{piu} on
a parameter $u$ is modified from a pure mixmaster dynamics, where $u$
varies over the discrete set $u_n$ determined by the corresponding dynamical
system, to a continuous deformation with $u>0$, with corresponding matrices
$$ \gamma(u) =\left( \begin{array}{cc} -u & 1 \\ 1 & 0 \end{array}\right) \in \GL_2(\R), $$
replacing the sequence of matrices $\gamma_n \in \GL_2(\Z)$ of \eqref{GL2Z}. The
action of $\gamma_n$ on the torus moduli, in either \eqref{thetatransf}
or \eqref{GL2Zact2}, then extends to a similarly defined action of the matrices
$\gamma(u)\in \GL_2(\R)$, which is no longer by isomorphisms of $\cA_\Theta$
(respectively, Morita equivalences of $\cA_{\theta_3}$) but by transformations
that deform $\cA_\Theta$ (respectively, $\cA_{\theta_3}$) to a family of
non-isomorphic $\cA_{\Theta(u)}$ (respectively, non-Morita 
equivalent $\cA_{\theta_3(u)}$).

\smallskip

We can check the effect of such a deformation of the Kasner parameters
in \eqref{piu}, with $u=u(t)$, on the Friedmann equation. We find 
$$ \frac{\dot{a}}{a} =  \frac{-u}{1+u+u^2}\, t^{-1} + \frac{\dot{u} \, (u^2-1)}{(1+u+u^2)^2}\,
\log t  $$
$$ \frac{\dot{b}}{b} =\frac{1+u}{1+u+u^2} \,t^{-1}- \frac{\dot{u} \, u \, (u+2)}
{(1+u+u^2)^2}\, \log t  $$
$$ \frac{\dot{c}}{c} =\frac{u(1+u)}{1+u+u^2}\, t^{-1} + \frac{\dot{u} \, (2u+1)}
{(1+u+u^2)^2}\, \log t , $$
so that the Hubble parameter
$$ H=\frac{\dot{\fa}}{\fa}=\frac{1}{3} ( \frac{\dot{a}}{a}+ \frac{\dot{b}}{b} + \frac{\dot{c}}{c})
= \frac{1}{3} t^{-1} $$
remains the same, along the deformation, as the Hubble parameter for the
initial mixmaster dynamics \eqref{Hubblemix}. In fact, the first terms in the above
expression for the logarithmic derivatives of the scale factors add up to
$t^{-1}$, while the second terms add up to zero.

\smallskip

Thus, through a deformation $u=u(t)$ of this sort, which modifies the Kasner exponents,  
one can disrupt the mixmaster dynamics and deform the underlying 
noncommutative $3$-torus $\cA_\Theta$ to a commutative torus, through 
the corresponding action on the torus moduli of the transformations
$\gamma(u)\in \GL_2(\R)$ as above, while maintaining the Friedmann 
equation unaffected.

\medskip
\subsection{Transition to slow-roll inflation}
As a second step, after interrupting the mixmaster oscillation and
damping the noncommutativity as described above, we want to
obtain a transition from the Friedmann equation for a Kasner metric
to the Friedmann equation for an inflationary cosmology. This can
also be obtained through a deformation, this time of the 
parameter $\beta$.

\smallskip

In fact, for a fixed plateaux value $\cV_\infty$ of the slow-roll potential $\cV(x)$
of \eqref{cVt}, we consider the expression
$$  \frac{\Lambda^4(t) \beta(t) \, \fa(t)^3}{4\pi} \cV_\infty.  $$
Here we may assume, as discussed in \S \ref{timesec}, that $\Lambda(t)\sim \fa(t)^{-1}$,
which leaves us with 
$$ \frac{\Lambda(t) \beta(t)}{4\pi} \cV_\infty \sim \frac{\beta(t)}{\fa(t)} \frac{\cV_\infty}{4\pi}. $$
A deformation from an initial phase with 
$$ \beta(t)\sim \frac{4\pi}{3\cV_\infty} \, \frac{\fa(t)}{t} $$
to a later phase with $\beta(t) \sim \fa(t)$ would then give the desired
transition from the Hubble parameter of a Kasner metric to the one of
a slow-roll inflationary cosmology.

\medskip

The two steps described in \S \ref{dampsec2} and in this subsection are
only a toy model, since we do not provide a viable physical mechanism
that produces the desired deformations affecting the transition to
commutativity and to the slow-roll inflation scenario. We only
exhibit a geometric model of how such transitions may be possible.

\bigskip
\section{Other aspects of the model}\label{OtherSec}

In this final section we outline briefly other aspects of the model of mixmaster
dynamics on noncommutative 3-tori, in relation to recently developed
methods for explicit computations of the spectral action \cite{ChCo};
to some existing computations, in the commutative case, of the modified
gravity terms in the asymptotic expansion of the spectral action for a
Kasner metric \cite{NeSa}; to the role of the diophantine condition
in the asymptotic expansion of the spectral action for noncommutative
tori \cite{EILS}, \cite{GIV}; and to the effect of coupling gravity to matter \cite{CMT}.
We also mention other physical models in which mixmaster dynamics
play an important role and for which the model presented in
this paper may have some relevance.

\subsection{The 4-dimensional Dirac operator for Euclidean Kasner metrics}

We now look at the 4-dimensional Euclidean version of the Kasner
metric, of the form $dt^2 + a(t)^2 dx^2 + b(t)^2 dy^2 + c(t)^2 dz^2$,
and in particular at the case where $a(t)=t^{p_1}$, $b(t)=t^{p_2}$, $c(t)=t^{p_3}$,
with $p_1+p_2+p_3=1=p_1^2+p_2^2+p_3^2$ as above. We adapt some
of the results obtained in \cite{ChCo} for the isotropic Euclidean
Robertson--Walker metrics to this non-isotropic case. 

\smallskip

One can consider the Euclidean version of the anisotropic metric \eqref{kasner},
of the form
\begin{equation}\label{Ekasner}
dt^2 +  a(t)^2 dx^2 + b(t)^2 dy^2 + c(t)^2 dz^2.
\end{equation}
One can proceed as in \cite{ChCo}, and construct the corresponding 4-dimensional
Dirac operator (with summation over repeated indices)
\begin{equation}\label{4dDirac}
\Dirac = \gamma^r e^\mu_r \frac{\partial}{\partial x^\mu} + \frac{1}{4} \gamma^s \omega_{srl} \gamma^{rl},
\end{equation}
where $\gamma^r$ are the gamma matrices, and the spin connection $\omega_{srl}$ now has nonzero terms
$$ \omega_{101}=\frac{\dot{a}}{a},  \ \ \ \  \omega_{202} = \frac{\dot{b}}{b}, \ \ \ \
\omega_{303} = \frac{\dot{c}}{c}. $$
Thus, one obtains an operator of the form
\begin{equation}\label{4dDirac2}
\Dirac = \gamma^0 (\frac{\partial}{\partial t} + \frac{3}{2} \frac{\dot{\fa}}{\fa}) + D,
\end{equation}
with $\fa$ the average scale factor as in \eqref{avscale}, and 
where the operator $D$ is given by
$$ D =  \gamma^1 \frac{1}{a} \frac{\partial}{\partial x} +  \gamma^2 \frac{1}{b} 
\frac{\partial}{\partial y} +  \gamma^3 \frac{1}{c} \frac{\partial}{\partial z}, $$
so that, as in the case of   \cite{ChCo}, one has $\gamma^0 D = -\dirac_{T^3_t} \oplus \dirac_{T^3_t}$, since $\gamma^0 \gamma^j = i\sigma_j$, with a sign difference due to our 
use of $-i \sigma_j$ instead of $i\sigma_j$
in \eqref{DiracT3}. We also have, as in \cite{ChCo}, that $\gamma^0 D = - D \gamma^0$
so that $D^2=(\gamma^0 D)^2$. 

It is convenient here to rename the operator $\dirac_{T^3_t}$ using the following
notation
\begin{equation}\label{Dabc}
\dirac_{a,b,c} : =  -i ( \sigma_1 \frac{1}{a} \frac{\partial}{\partial x} + \sigma_2 \frac{1}{b} 
\frac{\partial}{\partial y} + \sigma_3 \frac{1}{c} \frac{\partial}{\partial z} ).
\end{equation}
One can write the square of the Dirac operator $\Dirac^2$ as 
\begin{equation}\label{4Diracsquare}
 - (\frac{\partial}{\partial t}+ \frac{3}{2} \frac{\dot{\fa}}{\fa})^2 +
\dirac_{a,b,c}^2 - \dirac_{\frac{a^2}{\dot{a}},\frac{b^2}{\dot{b}},\frac{c^2}{\dot{c}}} ,
\end{equation}
since the cross term contributes only a term of the form $ \gamma^0 \frac{\partial}{\partial t} D + D \gamma^0 \frac{\partial}{\partial t} $ which gives 
$$ - i (\sigma_1 \frac{\partial}{\partial t}(\frac{1}{a}) \frac{\partial}{\partial x} + \sigma_2  \frac{\partial}{\partial t}(\frac{1}{b} )
\frac{\partial}{\partial y} + \sigma_3  \frac{\partial}{\partial t}(\frac{1}{c}) \frac{\partial}{\partial z} ) , $$
and, as above, $\gamma^0 D = -\dirac_{a,b,c} \oplus \dirac_{a,b,c}$.

Thus, one can reduce the spectral problem for this operator to an infinite family of one-dimensional problems, similarly to what happens in \cite{ChCo}. Here one uses the fact that
the two operators $\dirac_{a,b,c}$ and $\dirac_{\frac{a^2}{\dot{a}},\frac{b^2}{\dot{b}},\frac{c^2}{\dot{c}}}$ are simultaneously diagonalized with spectra
\begin{equation}\label{Spdiracabc}
\Spec (\dirac_{a,b,c})=\{ \pm 2\pi \| (\frac{k+\fs_1/2}{a}, \frac{m+\fs_2/2}{b}, \frac{n+\fs_3/2}{c}) \|\,|\, (k,m,n)\in \Z^3 \},
\end{equation}
\begin{equation}\label{Spdiracabc2}
\Spec (\dirac_{\frac{a^2}{\dot{a}},\frac{b^2}{\dot{b}},\frac{c^2}{\dot{c}}})=\{ \pm 2\pi \| (\frac{\dot{a}(k+\fs_1/2)}{a^2}, \frac{\dot{b}(m+\fs_2/2)}{b^2}, \frac{\dot{c}(n+\fs_3/2)}{c^2}) \|\,|\, (k,m,n)\in \Z^3 \}.
\end{equation}
We obtain in this way the one-dimensional problems
\begin{equation}\label{1dimprobl}
-(\frac{\partial}{\partial t}+ \frac{3}{2} \frac{\dot{\fa}}{\fa})^2 + \lambda_{a,b,c}^2 - \lambda_{\frac{a^2}{\dot{a}},\frac{b^2}{\dot{b}},\frac{c^2}{\dot{c}}},
\end{equation}
where we write $\lambda_{a,b,c}$ for the elements in the spectrum \eqref{Spdiracabc}
and $\lambda_{\frac{a^2}{\dot{a}},\frac{b^2}{\dot{b}},\frac{c^2}{\dot{c}}}$ for those in the
spectrum \eqref{Spdiracabc2}. This family \eqref{1dimprobl} of one-dimensional 
problems is then, in principle, suitable for a computation of the spectral action 
based on the Feynman--Kac formula, as described in \cite{ChCo}.

\smallskip

In the mixmaster case, one has spectra
$$ \lambda_{a,b,c}^2 =t^{-2p_1}(k+\frac{\fs_1}{2})^2 + t^{-2p_2} (m+\frac{\fs_2}{2})^2
+ t^{-2p_3} (n+ \frac{\fs_3}{2})^2 $$
$$ \lambda_{\frac{a^2}{\dot{a}},\frac{b^2}{\dot{b}},\frac{c^2}{\dot{c}}}^2 
= p_1^2 t^{-2(p_1+1)}(k+\frac{\fs_1}{2})^2 + p_2^2 t^{-2(p_2+1)} (m+\frac{\fs_2}{2})^2
+ p_3^2 t^{-2(p_3+1)} (n+ \frac{\fs_3}{2})^2, $$
since $a(t)=t^{p_1}$, $b(t)=t^{p_2}$, $c(t)=t^{p_3}$, with $\dot{a}/a^2= p_1 t^{p_1-1} t^{-2p_1}=
p_1 t^{-(p_1+1)}$ etc, while the first term in the operators \eqref{1dimprobl} of the 
one-dimensional problems takes the form
$$ - (\frac{\partial}{\partial t} + \frac{1}{2t} )^2, $$
since $\fa=t^{1/3}$ and $\dot{\fa}/\fa =t^{-1}/3$.

\medskip
\subsection{Asymptotic expansion and the Kasner metrics}

In the asymptotic expansion of the spectral action over a commutative
or almost-commutative geometry, one finds gravitational terms of the
form (see \cite{CCM})
$$ \int \left(\frac{1}{2\kappa_0^2} R + \frac{\alpha_0}{2} C_{\mu\nu\rho\sigma} C^{\mu\nu\rho\sigma}+\tau_0 R^*R^* - \xi_0 R |H|^2\right) \, dv, $$
where $C_{\mu\nu\rho\sigma}$ is the Weyl curvature, $R^*R^*$ is the form representative
of the Pontrjagin class, which integrates to the Euler characteristic, and $H$ is the Higgs field,
with the coefficients expressed in terms of the momenta of the test function $f$ in
the spectral action and the Yukawa coupling matrix for the matter part of the model.

\smallskip

An analysis of the gravitational terms in the asymptotic expansion
of the spectral action in the case of a Kasner metric (and more
generally for Bianchi V models) was given in \cite{NeSa}. In that
paper, the authors focus on the modified gravity term given by
the Weyl curvature that appears in the asymptotic expansion
of the spectral action, neglecting the term with the conformal
coupling of the Higgs field to gravity. They obtain modified
Einstein equations of the form
$$ R^{\mu\nu} -\frac{1}{2} g^{\mu\nu} R - \alpha_0 \kappa_0^2 
\left( 2 C^{\mu\lambda\nu\kappa}_{;\lambda;\kappa} - C^{\mu\lambda\nu\kappa} R_{\lambda\kappa} \right)= \kappa_0^2 T^{\mu\nu}_{matter}. $$

\smallskip

They compute this explicitly in the case of Kasner metrics and
other anisotropic Bianchi models, and they find that for such models
the obtained equations indeed differ from the ordinary Einstein--Hilbert
case, hence distinguishing the modified gravity model given by the
spectral action from ordinary GR.

\smallskip

In the mixmaster case, with ordinary commutative tori $T^3_t$, one can
see that their computation gives a correction term in the above equation 
with respect to the usual Einstein--Hilbert case of the form
$$ \frac{-4\alpha_0 \kappa_0^2}{3 t^4} \sum_i p_i \Big( p_1p_2p_3 +
p_{i+1}\Big( (p_i-p_{i+1})^2 - p_ip_{i+1})  $$ $$ + (p_i-1) \Big(\frac{1}{2} p_{i+1} (p_{i+1}-1) +
\frac{1}{2} p_{i+2} (p_{i+2}-1) - p_i (p_i-1)\Big) $$
$$ + (p_i^2+2-3 p_i+ (1-p_i)(p_i-p_{i+1}-p_{i+2})\Big)
(2p_i-p_{i+1}-p_{i+2})\Big). $$
As observed in \cite{NeSa}, this contribution is relevant in the early universe and
becomes negligible for later times.

\medskip
\subsection{The asymptotic expansion and the diophantine condition}

In the case where the ordinary tori $T^3_t$ in the mixmaster model are
replaced by noncommutative tori $\bT^3_\Theta$, dealing with the 
asymptotic expansion of the spectral action becomes more delicate.

\smallskip

The perturbative expansion of the spectral action for noncommutative
tori was computed recently in \cite{EILS}, \cite{GIV}. In these calculations,
one considers the spectral action 
\begin{equation}\label{spectralDA}
\Tr( f(\dirac^2_A/\Lambda^2)), \ \ \ \text{ where } \ \ \  \dirac_A = \dirac + \tilde A, \ \ \ \text{ with } \ \ \  
\tilde A = A + \epsilon J A J^{-1},
\end{equation}
where $\epsilon$ is the commutation sign $J \dirac = \epsilon \dirac J$ of the real
structure involution $J$ and the Dirac operator and $A$ is a self-adjoint one form
$A=\sum_i a_i [\dirac, b_i]$, with $a_i$ and $b_i$ in $\cA_\Theta$. While, as we 
recalled in the previous sections,  the Dirac operator $\dirac$ on $\bT_\Theta$ is 
inherited via isospectral deformation from a commutative torus, the operator $\dirac_A$,
twisted with the inner fluctuations $\tilde A$ as above, depends on the noncommutative
torus itself, and the results of  \cite{EILS}, \cite{GIV} show that this dependence
manifests itself in an explicit dependence of the spectral action $\Tr( f(\dirac^2_A/\Lambda^2))$
on refined number-theoretic information on the modulus $\Theta$ of the noncommutative
torus. The computation of the spectral action in \cite{EILS} is through the
asymptotic expansion for large $\Lambda$, in the form 
\begin{equation}\label{spactAexpand}
\Tr( f(\dirac^2_A/\Lambda^2)) = \sum_{k\in \dim\Spec^+} f_k \, \Lambda^k \, \cutint |\dirac_A|^{-k} 
+ f(0) \, \zeta_{\dirac_A}(0) + O(\Lambda^{-1}),
\end{equation}
where the sum is over points in the strictly positive part of the dimension spectrum
and the integration
$\cutint |\dirac_A|^{-k}$ can be expressed as a residue of the zeta function of $\dirac_A$ 
at the pole $s=k$.

In particular, a diophantine condition on $\Theta$ plays a crucial role in the results
of \cite{EILS}, \cite{GIV}. This ``badly approximable" condition is formulated as follows. 
A vector $\underline{v}$ in $\R^n$ 
is $\delta$-diophantine, for some $\delta>0$, if there exists a $C>0$ so that 
\begin{equation}\label{badapprox}
 |\underline{v}\cdot \underline{q} - 2\pi k| \geq C\, |\underline{q}|^{-\delta}, \ \ \ \  \forall q \in \Z^n
 \backslash \{0\}, \ \  \forall k \in \Z.
\end{equation}
The set of diophantine vectors is the union over $\delta$ of the sets of
$\delta$-diophantine vectors. A matrix $\Theta$ is diophantine if there is
a vector $\underline{v}\in \Z^n$ such that $\Theta \underline{v}$ is diophantine.
Almost every matrix is diophantine (with respect to the Lebesgue measure).

\medskip

The positive dimension spectrum for an $n$-dimensional noncommutative torus 
$\bT^n_\Theta$ consists of the points $\{ 1,2,\ldots, n \}$, which are all simple poles,
under the assumption that the matrix $\frac{1}{2\pi} \Theta$ is diophantine.  Under
this same diophantine assumption, the residues at the points of the dimension
spectrum are computed explicitly in \cite{EILS}. The top $n$-th term is the same as for
the unperturbed Dirac operator $\dirac$; the $(n-k)$ terms with $k$ odd vanish, and the
$(n-2)$ term also vanishes. 

\medskip

Thus, in the case of a 3-torus, the only nonvanishing term agrees with the 
unperturbed case, so that the resulting perturbative spectral action under 
the diophantine condition does not differ in form from the ordinary torus case.
By comparison, the case of a $4$-dimensional torus exhibits a more interesting 
phenomenon whereby the perturbative spectral action in the case satisfying
the diophantine condition has the form  
$$  8\pi^2 f_4 \Lambda^4 - \frac{4\pi^2}{3} f(0)\tau(F^{\mu\nu}F_{\mu\nu}) + O(\Lambda^{-2}) ,$$
where $F^{\mu\nu}$ is while for the commutative case one would have $\dirac_A=\dirac$, since the inner
fluctuations would all be equivalent to zero in that case, so the expansion under the
diophantine condition is different in this case.

\medskip

The transformation \eqref{thetatransf} of the parameters $\underline{\theta}$ 
in the mixmaster dynamics via the matrix \eqref{GL2Z} preserves the diophantine
condition, hence in a phase of pure mixmaster dynamics, the asymptotic
expansion of the spectral action of the noncommutative tori given in 
\cite{EILS}, \cite{GIV} remains valid. However, in the presence of a damping
effect that destroys the mixmaster evolution and transitions to an isotropic 
and commutative torus geometry, the parameters $\underline{\theta}$ are
deformed to zero along a transformation that no longer preserves the
diophantine condition. In fact, along such a deformation, the parameters
$\underline{\theta}$ will hit infinitely many values that do and do not satisfy 
the Diophantine condition and the properties of the asymptotic
expansion of the spectral action may vary accordingly in a seemingly chaotic
manner, until the final stage of commutative isotropic tori is reached. 
Even though for $3$-tori, unlike the case of $4$-tori, the unperturbed commutative
spectral action ends up looking the same, one should still investigate what
happens in the intermediate stages of the evolution of the parameter $\underline{\theta}$
in between an initial anisotropic mixmaster noncommutative case satisfying
the diophantine condition and a final isotropic commutative case, especially 
at all the intermediate values of $\underline{\theta}$ that do not satisfy the
diophantine condition.

\smallskip
\subsection{Coupling to matter}

In addition to pure gravity, described by the spectral action on the
mixmaster tori (commutative or noncommutative), one can have
a nontrivial coupling to matter, by taking the product of the spectral
triple with a finite noncommutative geometry $F$, which specifies
the matter content of the model, as in \cite{CCM} and \cite{CoSM}.

\smallskip

In terms of the asymptotic expansion of the spectral action, we can follow 
the computations of \cite{GIV}. We now consider the example of a product
geometry of a noncommutative torus $\bT^n_\Theta$ and a finite geometry
of the form 
\begin{equation}\label{Fgeom}
\cA_F = M_q(\C), \ \ \  \cH_F = M_q(\C),  \ \ \ \  D_F =0, \ \ \  J_F =J_q,
\end{equation}
where $J_q(L(T))=L(T^*)$, for $T$ an element in the algebra and 
$L(T)$ its representation on the Hilbert space. 

\smallskip

It is also convenient to write the spectral geometry for the noncommutative
torus in the form $\cH=\cH_\tau$, the GNS representation for the
tracial state $\tau(a)=a_0$ on $\cA^\infty_\Theta$, so that, $\tau$ being faithful,
this is the Hilbert space completion of $\cA^\infty_\Theta$ in the inner
product $\langle a, b \rangle = \tau(a^* b)$. Then the real structure is $J_0: \cH_\tau \to \cH_\tau$,
$J_0(h)=h^*$, with $J_0^{-1}=J_0$ and, for $L(a)$ an element of $\cA^\infty_\Theta$ acting
by left multiplication on $\cH_\tau$,
$$ J_0L(a)J_0^{-1}h = J_0L(a)J_0h = J_0L(a)h^* = hL(a^*)\equiv R(a^*)h,$$ 
where $R(a)$ is the action of $a$ by right multiplication. Then the spectral geometry
from the noncommutative torus can be written as
$$ \cA=\cA^\infty_\Theta, \ \ \  \cH =\cH_\tau \otimes \C^{2^{\lfloor n/2\rfloor}}, \ \ \ 
\dirac =  i\delta_\mu \otimes\gamma^\mu, \ \ \  J=J_0\otimes C_0, $$
where $\mathbb{C}^{2^{\lfloor n/2\rfloor}}$ is an irreducible representation of $\mathbb{C}l^{(+)}(\mathbb{R}^n),$ the $\gamma^\mu$ are the gamma matrices implementing the representation, and $C_0$ is the antiunitary operator on $\mathbb{C}^{2^{\lfloor n/2\rfloor}}$ implementing the charge conjugation.  The $\delta_\mu$ are the basic derivations on the noncommutative torus,
acting as
$$\delta_\mu\left(\sum_r{a_ru^r}\right)\equiv i\sum_r{r_\mu a_ru^r},$$
on elements of $\cH_\tau$ written as $a = \sum_r{a_ru^r}$, where
$r\in\mathbb{Z}^n,$ and $\{a_r\}\in\mathcal{S}(\mathbb{Z}^n),$ and $u^r$ is a 
Weyl element in $\cA^\infty_\Theta$ defined by
$$u^r := \exp\left[\pi i\sum_{j<k}{r_j\theta_{jk}r_k}\right] u_1^{r_1}u_2^{r_2}...u_n^{r_n}.$$

The sign of the spectral triple is 
$1\otimes\gamma^{n+1}$ for $n$ even. One has 
the relations $J\mathcal{D} = \epsilon\mathcal{D}J,$ and $J^2 = \epsilon' = (-1)^{\lfloor n/2\rfloor}.$  The first implies that $C_0\gamma^\mu = -\epsilon\gamma^\mu C_0.$

As shown in \cite{Ven}, this way of writing the spectral triple structure on the noncommutative
torus is equivalent to the one we used above by isospectral deformation.

\smallskip

The product geometry then has algebra $\cA^\infty_\Theta \otimes M_q(\C)$,
Hilbert space $\cH \otimes M_q(\C)=\cH_\tau \otimes \mathbb{C}^{2^{\lfloor n/2\rfloor}} \otimes M_q(\mathbb{C})$, Dirac operator $D= i\delta_\mu \otimes\gamma^\mu \otimes \textrm{id}_q$
and real structure $J=J_0\otimes C_0\otimes J_q$.

\smallskip

Then, following the same computation of \cite{GIV}, one finds that the presence of
the finite spectral triple $F$ of \eqref{Fgeom}
alters the perturbative expansion of the spectral action 
by a factor of $q^2$ in the leading order term. For example, in the case of a $4$-dimensional
torus, under the assumption that the diophantine condition on $\Theta$ holds, one
finds 
$$ 8\pi^2 q^2 f_4 \Lambda^4 - \frac{4\pi^2}{3} f(0)\tau(F^{\mu\nu}F_{\mu\nu}) 
+ O(\Lambda^{-2}) .$$

\smallskip

In the case of the commutative mixmaster tori $T^3_t$, one can
also proceed as in \cite{CMT} to compute the effect on the gravitational
part of the spectral action of the matter sector, which as in \cite{CMT} also delivers 
an overall multiplicative factor $q^2$, equal to the rank of the finite geometry.

\smallskip

Thus, in the case of the mixmaster tori, switching back to the same
notation used in the previous sections, we find that the spectral action,
for this case with the finite geometry $F$ of \eqref{Fgeom}, 
\begin{equation}\label{spactT3averMq}
\Tr( f(D_t^2/\Lambda^2)) \sim
\frac{q^2 \fa(t)^3 \Lambda^3}{4\pi^3} \int_{\R^3} f(u^2+v^2+w^2)\, du\, dv\, dw \, .
\end{equation} 
Correspondingly, when one computes the inflation potential as in \eqref{TrhDphi},
one finds that, as in the case of \cite{CMT}, the resulting slow roll potential 
is $q^2 V(x)$, with $V(x)$ the inflation potential in the absence of the finite geometry.
Since only the amplitude of the potential is affected, the slow-roll parameters
do not change, but as argued in \cite{MPT2} and \cite{CMT}, 
the amplitudes for the power spectra for density perturbations and gravitational waves 
(scalar and tensor perturbations) detect the different scaling factors in the slow-roll 
potentials and can therefore, in principle, detect the presence of the matter sector
through the rank of the finite geometry.

\medskip
\subsection{Relation to other physical models}

We mention here briefly other recent cosmological models where
the kind of noncommutative deformation of the mixmaster dynamics
we described above may turn out to be useful.

\medskip

A first possible context is Ho\v{r}ava--Lifschitz gravity.
It was recently shown in \cite{BBLP} that the mixmaster universe
provides a mini-superspace truncation of the field equations of
Ho\v{r}ava--Lifschitz gravity. The latter is a recently introduced 
higher derivatives modified gravity theory, applicable in the 
ultraviolet regime \cite{Hor}. The Lagrangian density for this
theory is derived from a superpotential that contains 
a Chern--Simons gravitational term and 
a 3-dimensional Einstein--Hilbert term. The resulting expression
has terms of the form
$\alpha R +\beta + \gamma \cC_{ij} \cC^{ij}
+\delta \cC_{ij}R^{ij} +\epsilon R_{ij} R^{ij} +\zeta R^2$,
where the first two terms give rise to the usual Einstein--Hilbert action and
the remaining higher derivative terms contain the Cotton tensor $\cC^{ij}$
(see (2.16) of \cite{BBLP}), which vanishes if the 3-dimensional spatial
sections are conformally flat, and the curvature tensors. The action is invariant
under coordinate transformations of the spatial sections.  

\medskip

Another such context is loop quantum cosmology. It was recently shown in \cite{Sloan}
that, in the setting of loop quantum cosmology, one finds an oscillatory behavior of
mixmaster type as one approaches the singularity as a simplified system from the
equations of motion in the Hamiltonian formulation (see \S 2.3--2.8  
in \cite{Sloan}) and that, moreover, on the resulting reduced phase space one can
also introduce a scalar field with an inflation slow-roll potential (see \S 2.7 and 
chapter 4 of \cite{Sloan}). The presence of the scalar field has a damping effect on
the mixmaster oscillations in this model (see Appendix A of \cite{Sloan}).

\subsection*{Acknowledgment} The first author was supported for this project by
a Caltech Summer Undergraduate Research Fellowship and by a Richter Scholarship,
provided by the Richter Memorial Fund. The second author
is supported by NSF grants DMS-0901221 and DMS-1007207.

\end{document}